# Ion Implantation Enhanced Nucleation Facilitates Heat Transport across Atomically-Sharp Semiconductor Interfaces

Jinwen Liu[1,+], Zifeng Huang[2,+], Lina Yang[3,+], Yachao Zhang[4], Xingqiang Zhang[2], Kun Zhang[4], Xufei Guo[2], Yuxiang Wang[2], Hong Zhou[4,*], Jincheng Zhang[4], Wei Wang[1,2], Yue Hao[4], Zhe Cheng[1,2,*]

1 School of Software & Microelectronics, Peking University, Beijing 100871, China

2 School of Integrated Circuits and Beijing Advanced Innovation Center for Integrated Circuits, Peking University, Beijing 100871, China

3 School of Aerospace Engineering, Beijing Institute of Technology, Beijing 100081, China

4 State Key Discipline Laboratory of Wide Bandgap Semiconductor Technology, School of Microelectronics, Xidian University, Xi'an 710071, China

[+] These authors contributed equally

[*]Authors to whom correspondence should be addressed: hongzhou@xidian.edu.cn; zhe.cheng@pku.edu.cn

**Keywords:** Thermal boundary conductance, phonon transport, electron energy loss spectroscopy, electronics cooling.

**Abstract**

Overheating is a critical bottleneck limiting the performance and reliability of next-generation high-power and high-frequency electronics. Interfacial thermal resistance constitutes a significant portion of the total thermal resistance. In this study, we report an ultrahigh thermal boundary conductance (TBC) of approximately 800 MW/m$^2$-K at the atomically-sharp AlN-SiC interface, achieved through an ion implantation-enhanced nucleation epitaxy technique. This value is among the highest TBC values reported for semiconductor interfaces, confirmed by structural characterizations which show an ultrahigh-quality interface. Atomistic Green's Function calculations reveal that elastic phonon transmission dominates the interface, with nearly half of the acoustic modes (0-15 THz) exhibiting near-unity transmission due to the atomically sharp structure. Furthermore, using high-energy-resolution electron energy loss spectroscopy, we probe vibrational properties with nanometer spatial resolution and identify unique interfacial phonon modes connecting the mismatched phonon spectra, confirmed by molecular dynamics simulations. The ultrahigh TBC is attributed to both the high elastic phonon transmission due to the high quality interfaces and the inelastic phonon scattering channel due to interfacial phonon modes. These findings not only advance the fundamental understanding of interfacial thermal transport but also provide a pathway for effective thermal management in emerging electronic devices.

**Introduction**

Wide-bandgap semiconductors such as gallium nitride (GaN) have demonstrated significant potential in power and radio-frequency (RF) electronics, fueling technological progress in applications including satellite communications, 5G infrastructure, and radar systems.[1–5] Yet, the thermal management of GaN-based high-electron-mobility transistors (HEMT) is crucial in determining their output power density and operational reliability. A mere 10 K rise in junction temperature cuts the lifetime of the GaN device in half.[6] In GaN devices, the nanoscale hot spot generates localized heat flux that can surpass the heat flux observed on the sun's surface by an order of magnitude[7]. Given the distinctive thermal behavior of GaN HEMTs—specifically their ultra-high heat flux and nanoscale heating zones—effective heat spreading near the junction becomes essential to lower the peak temperature.

Silicon carbide (SiC) is widely used as a substrate material for GaN RF devices. Nevertheless, direct epitaxial growth of GaN films on SiC leads to low-quality GaN crystal, primarily due to lattice mismatch between the two materials. Aluminum nitride (AlN) is recognized for its high thermal conductivity (reaching up to 321 W/m-K) and excellent electrical insulation properties[6,8]. Its lattice constant and coefficient of thermal expansion (CTE) are also well matched with both GaN and SiC, making it a promising transition layer.[9,10]

The spreading of heat from the localized hotspot is strongly affected by the proximity of the AlN-SiC interface, rendering interfacial thermal engineering a pivotal design aspect.[10] Recent studies on thermal transport across AlN-SiC interfaces grown by conventional techniques have revealed inconsistent TBC values, including 310 MW/m$^2$-K for an epitaxial AlN-3C-SiC interface,[1] and 470 MW/m$^2$-K for an epitaxial AlN-4H-SiC interface.[9] Transmission electron microscopy (TEM) analyses suggest that these variations arise from substantial interfacial imperfections that intensify phonon scattering. Experimental data obtained via the 3ω method and frequency-domain thermoreflectance (FDTR) show a TBC value of 196 MW/m$^2$-K for metal-organic vapor phase epitaxy (MOVPE)-grown AlN-SiC interfaces.[11] In comparison, non-equilibrium molecular dynamics (NEMD) simulations predicted a low TBC of 111 MW/m$^2$-K.[12] A multiscale modeling approach combining non-equilibrium Green's function with density functional theory (DFT)[13] achieved consistency with the experimental TBC value only when high-density oxygen defect complexes were incorporated at the interface. The preceding analysis demonstrates that interfacial defect between AlN and SiC critically governs thermal transport, necessitating atomically sharp interfaces to achieve high TBC. However, a unified structural-thermal property relationship between interfacial structure and TBC is still missing.

In this work, we use an ion implantation technology to enhance the nucleation of AlN when epitaxially growing on SiC substrate to form high-quality interfaces. High-resolution X-ray diffraction (HR-XRD) and atomic-scale scanning transmission electron microscope (STEM) are used to characterize the interfacial structures.

Furthermore, time-domain thermoreflectance (TDTR) is performed to measure the TBC of the AlN-SiC interfaces. Finally, to understand the structure-TBC relationship, high-energy-resolution electron energy loss spectroscopy (EELS) is utilized to map the vibrational properties across the interface with nanometer spatial resolution, corroborated by MD simulations.

**Results**

By using ion implantation enhanced nucleation technology, a layer of AlN with a thickness of 430 nm is epitaxially grown on SiC substrates (Figure 1a), unlike conventional growth methods which typically involve depositing island-like AlN directly onto silicon carbide (SiC) substrates. Selective etching is applied to achieve three distinct AlN thicknesses (117 nm, 241 nm, 337 nm). Spectroscopic ellipsometry and cross-sectional STEM (see Figure S6) are used to further confirm the thickness of the epitaxial AlN layer. X-ray diffraction measurements were performed to assess the structural properties of the AlN films hetero-epitaxially grown on the 4H-SiC substrates. Peaks for AlN (0002) and ($10\bar{1}2$) are observed. Figure 1b shows that the ω-scan of the (0002) reflection exhibits a dual-peak profile (17.83° and 18.02°), which is from the AlN (0002) crystal plane and the 4H-SiC substrate (0004) crystal surface.[14–16]

The crystalline quality was further evaluated through off-axis reflections. Figure 1c shows the ($10\bar{1}2$) reflection, which exhibits a sharp diffraction peak at 24.90° with an ultra-narrow full width at half maximum (FWHM) of 54 arcseconds. This low FWHM value signifies a very low density of extended defects and underscores the high quality of the epilayer.[17] The corresponding edge-type dislocation density was calculated to be $1.63\times10^7$ cm$^{-2}$,[14,18] which is consistent with the notion that the FWHM of such asymmetric rocking curves is influenced by both tilt and twist.[19,20] Furthermore, the in-plane alignment was quantified by a φ-scan of the $(10\bar{1}2)$ plane (Figure 1d). The scan demonstrates a clear sixfold symmetry, with peaks separated by nearly 60°±0.1°. All peaks have narrow FWHM values between 0.023° and 0.050° (mean: 0.038°), confirming minimal twist dispersion. To conclude, the XRD results show the single-crystal epitaxy of AlN films with high orientational uniformity.

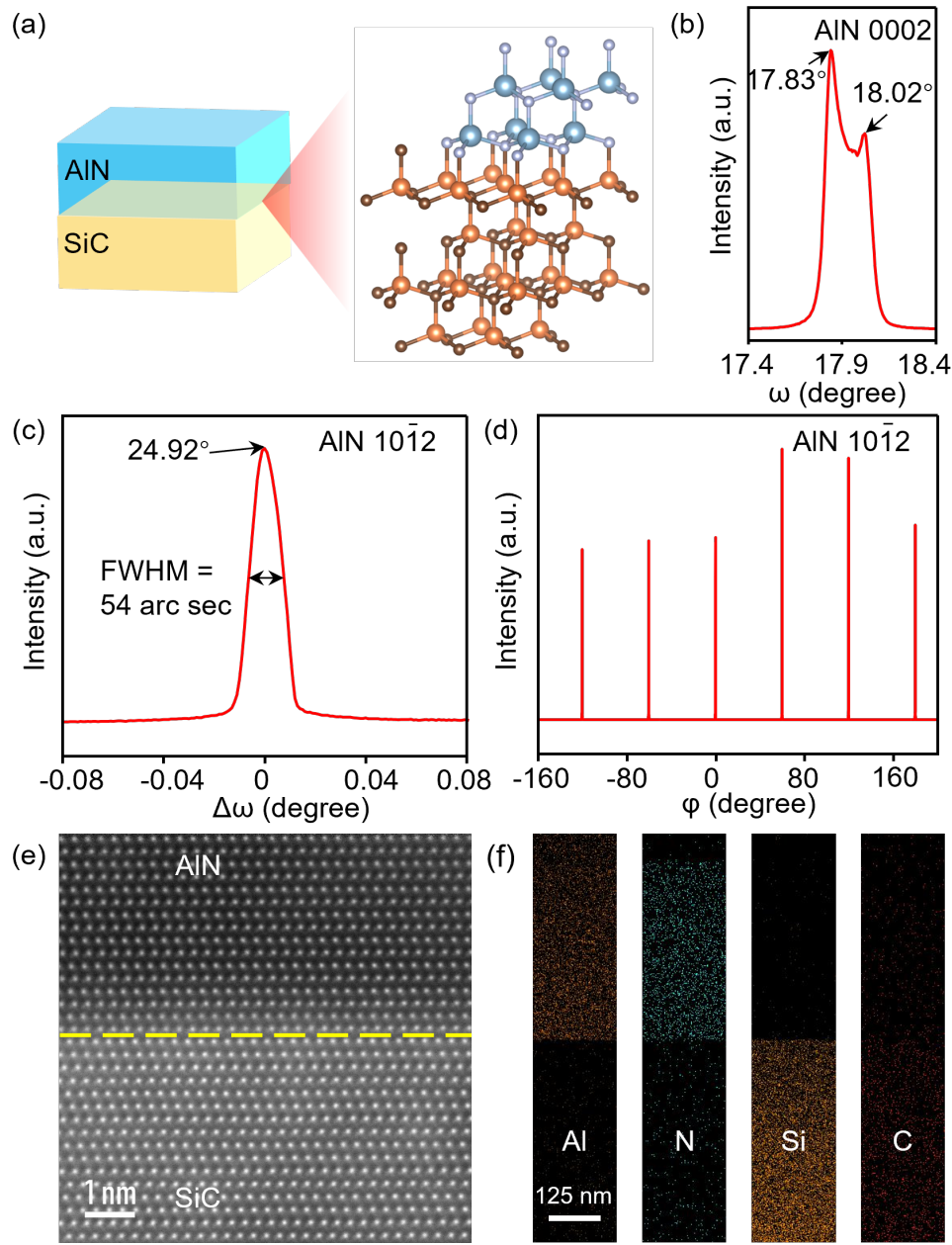

Figure 1. Structural characterization of the AlN thin film. a) Atomic structure diagram of the AlN-SiC interface. b) 2θ-ω scan of the AlN (0002) crystal plane, showing a dual-peak which corresponds to AlN and SiC. c) Rocking curve for the AlN ($10\bar{1}2$) crystal plane, exhibiting a narrow peak. d) φ-scan for the AlN ($10\bar{1}2$) crystal plane. e) Atomic-resolution image of the AlN-SiC junction. f) Elemental mapping using Energy-dispersive X-ray spectroscopy (EDS) across the interface.

The microstructure of the AlN epitaxial layer grown on the SiC substrate were further elucidated using high-resolution scanning transmission electron microscopy (HR-STEM). Figure 1e shows that the AlN-SiC interface is atomically sharp, and the EDS analysis in Figure 1f confirms that interfacial mixing is negligible. The high crystallinity of the AlN epilayer is evidenced by the high-resolution lattice image in Figure S7. Recent theoretical studies indicate that, for lattice-matched systems, interfacial mixing introduces mass disorder that acts as phonon scattering sources, reducing phonon transmission.[21,22] Therefore, the absence of mixing at the interface in this work is favorable to enhance phonon transmission, thereby enhancing efficient thermal transport across interfaces.

To characterize the thermal properties, we employed TDTR on a series of AlN-SiC samples with varying AlN layer thicknesses to extract the thermal conductivity of both the AlN epitaxial layer and the SiC substrate, as well as the TBC at their interfaces. The experimental setup is depicted in Figure 2a (see Methods for more details). Each specimen was coated with an approximately 80 nm aluminum (Al) film. Our procedure started with measuring a bare 4H-SiC substrate to determine its thermal conductivity, which served as a fixed parameter when analyzing the AlN-SiC samples. Subsequently, the AlN layer thermal conductivity and the AlN-SiC TBC were determined for samples with different AlN thicknesses. A representative fitting curve of the TDTR signals is illustrated in Figure 2b.

As presented in Figure 2c, the thermal conductivity of the AlN films follows a power-law decay ($k = a \cdot T^{-b}$) with rising temperature. The fitted exponent $b$ for bulk AlN is ~1.4.[22,23] Smaller values are observed for the thinner films (from 430 nm to 117 nm). This trend arises from the distinct scattering mechanisms involved. Bulk material is governed solely by phonon-phonon scattering, which intensifies at higher temperatures and leads to a strong thermal conductivity reduction. As the AlN thickness approaches the scale of the phonon mean free path (MFP), however, boundary scattering becomes significant. Since such scatterings are temperature-independent, it mitigates the thermal conductivity decline in thinner films. The measured thermal conductivity are close to the theoretical values of perfect single crystal AlN thin films with similar thicknesses, indicating the high quality of the AlN thin films. As shown in Figure 2d, the TBC of our AlN-SiC interface reaches ~800 MW/m²-K at room temperature. This result not only surpasses all previously experimental reports for similar interfaces[1,9,11,12] but also ranks among the highest values documented for any semiconductor interfaces (see Figure 2e). Using a kinetic-gas upper-bound estimate ($\tau = 1$), our measured ~800 MW/m²-K corresponds to ~40% of the theoretical maximum TBC (see Section Max TBC in the SI). The high quality of the interface sets a new benchmark for this material system.

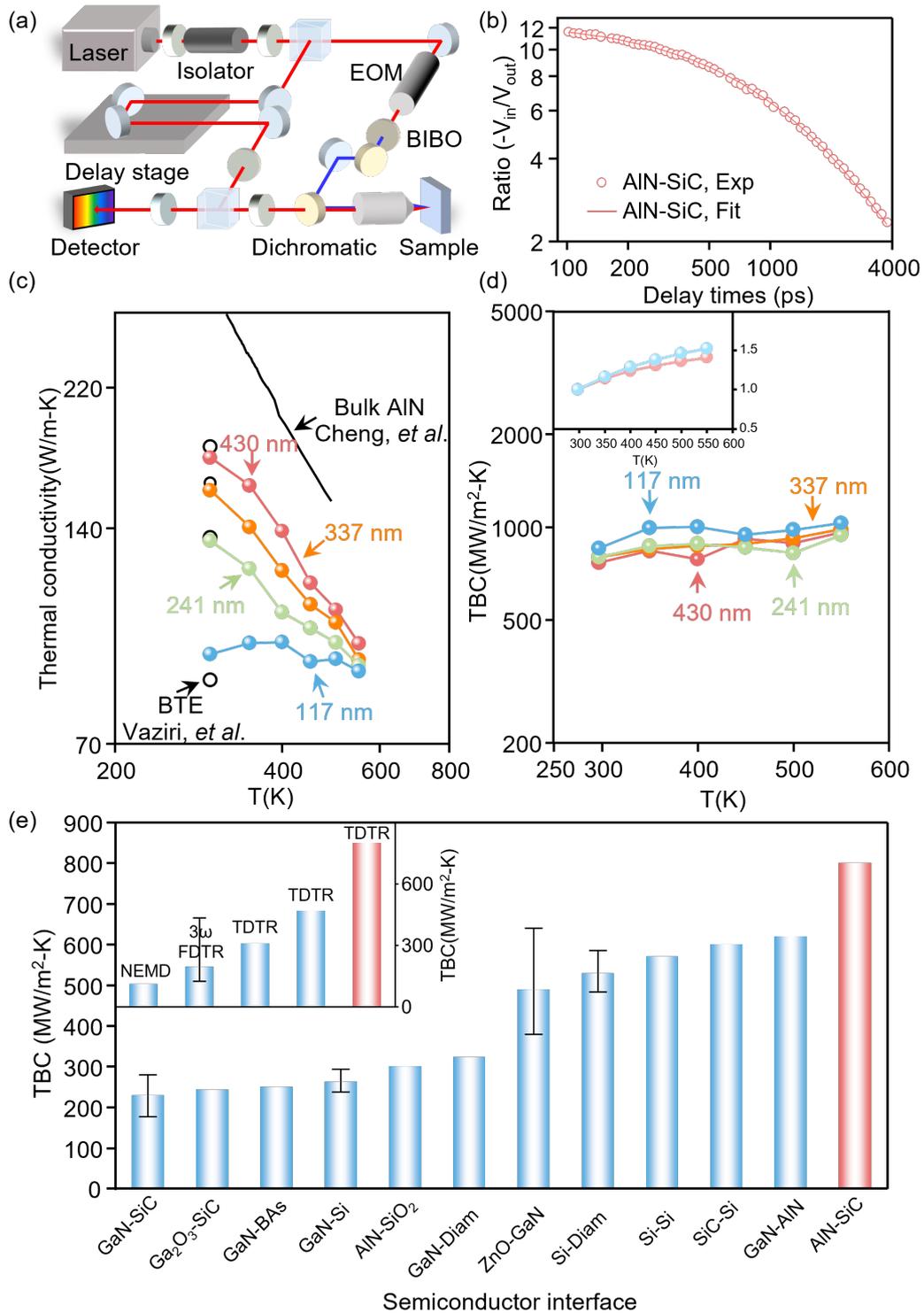

Figure 2. Thermal characterizations of the AlN-SiC samples by TDTR. a) Schematic diagram of the two-color TDTR setup. b) Representative data fitting of a TDTR measurement: experimental data (circles) and corresponding model fitting line (solid line). c) Thermal conductivity of AlN films plotted as a function of thickness, compared

with calculated bulk AlN data from Cheng *et al*.[22] d) The measured AlN-SiC TBC as a function of temperature. Inset: temperature-dependent trends of the normalized specific heat capacities of AlN (pink line) and SiC (blue line). e) TBC of the AlN-SiC interface benchmarked against other semiconductor interfaces, including GaN-SiC,[24] $Ga_2O_3$-SiC,[25] GaN-Bas,[26] GaN-Si,[27] AlN-$SiO_2$,[8] GaN-Diamond,[28] ZnO-GaN,[29] Si-Diamond,[30] Si-Si,[31] SiC-Si,[1] and GaN-AlN.[32] Inset: comparison of our measured TBC values with those reported in the literature.[1,9,11,12]

The temperature-dependent TBC shown in Figure 2d provides insights into the interfacial thermal transport. As detailed in the inset, we compare the measured TBC (normalized to 297 K) with the normalized volumetric heat capacities of bulk AlN and SiC. The TBC exhibits a weak monotonic increase from 297 K to 550 K, a trend that is consistent with the temperature dependence of the heat capacities of the constituent materials. The TBC is expected to scale with the specific heat if the phonon transmission remains stable across the temperature range. Therefore, the observed similarity in the increasing trends suggests that the TBC at elevated temperatures is primarily driven by the increased population of phonon modes.[33]

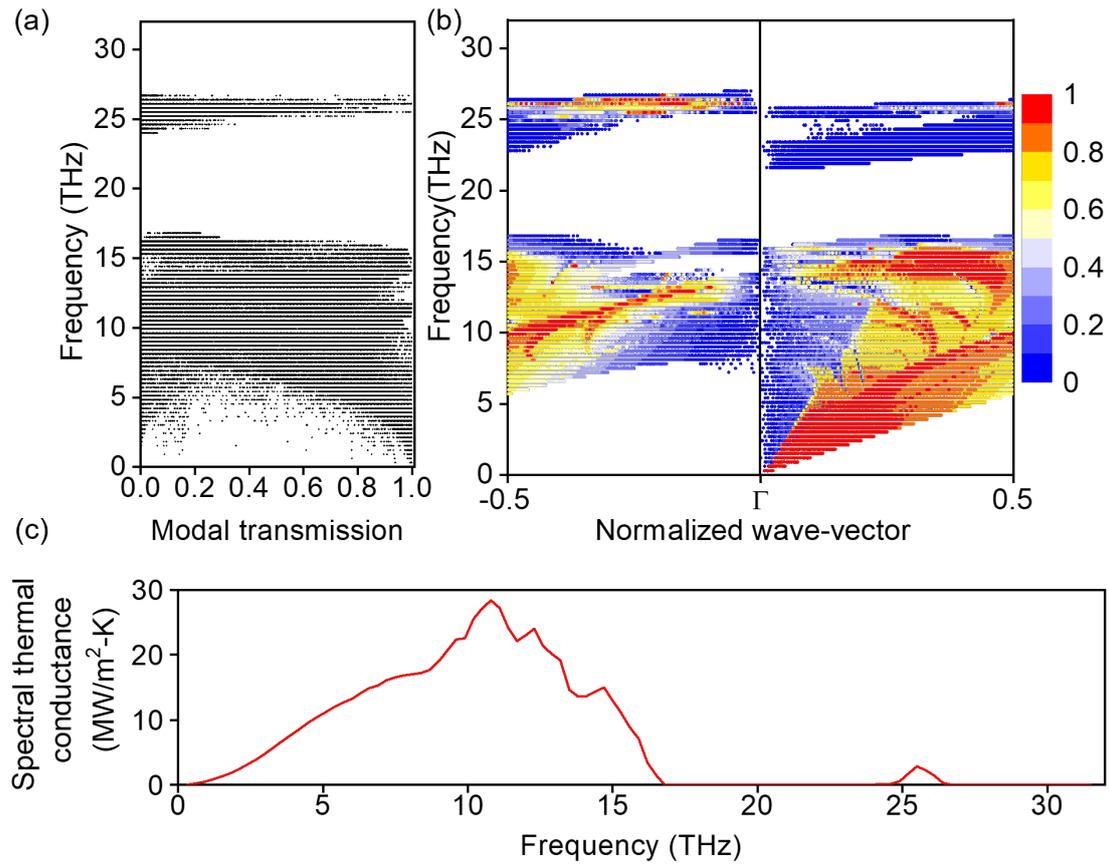

Figure 3. AGF-calculated phonon transport properties across the AlN-SiC interface. (a) Modal transmission showing near-unity transmission for nearly half of the acoustic modes and high transmission (> 0.5) for optical modes up to ~17 THz and near 25 THz. (b) Momentum-resolved transmission spectrum, where the red acoustic branches across the whole Brillouin zone indicate dominant near-unity transmission. The color scale (right) indicates the modal transmission ranging from 0 (blue) to 1 (red). (c) Spectral thermal conductance at 300 K. The integrated theoretical TBC (~735 MW/m$^2$-K) accounts for >90% of the experimental value, confirming that elastic transmission dominates the heat transport.

To quantitatively decouple the contribution of elastic phonon transport from the total

TBC, we performed the extended Atomistic Green's Function (AGF) calculations based on the atomically matched interface structure,[34,35] the details of settings are in section of Methods. Figure 3a displays the modal transmission distribution with respect to phonon frequency. In the low-frequency regime, nearly half of the acoustic modes exhibit near-unity transmission. Our AlN-SiC interface sustains high modal transmission (> 0.8) up to ~17 THz. Previous studies have further shown that, for sharp AlN-SiC interfaces, the frequency-resolved interfacial transport is predominantly concentrated below ~20 THz.[36] Moreover, appreciable transmission persists for high-frequency optical modes in the 24-27 THz range. This indicates that the excellent lattice matching not only facilitates acoustic phonon transport but also enables strong transmission of optical modes between AlN and SiC.

The underlying mechanism is further visualized in the momentum-resolved transmission spectrum shown in Figure 3b. The vivid red color of the acoustic branches throughout the entire first Brillouin zone signifies a near-unity transmission characteristic.[34] This implies that the interface is atomically smooth enough to conserve the phonon momentum, allowing heat-carrying phonons to traverse the junction with minimal diffuse scattering. These features establish the microscopic basis for elastic heat transfer across an atomically sharp interface.

To quantify the contribution of these transmission channels to the total interfacial heat transport, we calculated the spectral thermal conductance at 300 K, as shown in Figure

3c. The profile reveals that the heat transport is dominated by phonons in the 0-15 THz range, with a distinct peak around 10-12 THz. By integrating the spectral conductance curve, we obtained a theoretical TBC of ~735 MW/m$^2$-K. Remarkably, this value—derived solely from elastic phonon transmission assumptions—accounts for over 90% of our experimentally measured TBC (~800 MW/m$^2$-K). Notably, this elastic prediction is close to our experimentally measured value. This quantitative consistency suggests that the ultrahigh TBC is predominantly driven by the efficient elastic transmission enabled by the atomically sharp interface and excellent lattice and phonon spectrum matching.

To further investigate the fundamental mechanism of the high TBC, the phonon properties across the interfacial region were examined using high-energy-resolution electron energy loss spectroscopy with a nanometer resolution. As shown in Figure 4a, the phonon energy of the bulk SiC, bulk AlN, and the interfacial region was clearly identified. The energies of the acoustic and optical phonon branches of 4H-SiC and AlN are in good agreement with reference calculations,[37,38] demonstrating the high accuracy of the experiment. A good match is observed between the low-frequency acoustic phonon modes of SiC and AlN, which enables efficient elastic phonon transfer through the interface.[39–41] This effect can also be observed in Figure 4b, which greatly enhances interfacial heat transfer, since low-energy acoustic phonons are the main heat carriers in semiconductors. In contrast, there is a mismatch between the high-frequency optical phonon spectra of bulk AlN and SiC, which hinders direct elastic phonon transport across the interface. Notably, the optical phonon energies of both bulk materials exhibit an obvious shift near the interface, marked by orange circles in Figure 4a, connecting the lattice vibrations of the two sides. This phenomenon is more clearly illustrated in Figure 4b, where the phonon energy spectra of three different sections are shown separately, and the shifts of the peak positions are marked by orange arrows. A blue shift of the AlN optical modes and a red shift of the SiC optical modes are clearly observed. In addition, the residual phonon spectrum around 70 meV connects the SiC acoustic modes and AlN optical modes, as marked by the light green circle in Figure 4a.[42] This feature results from the broadening of the optical phonon peak in the interfacial region, as indicated by the light green arrow in Figure 4b. In the high-energy region (110 meV - 130 meV), we observed a phonon signal localized at the interfacial region, extending beyond the phonon energy ranges of both SiC and AlN, as indicated in Figure 4b and shown in the right panel.

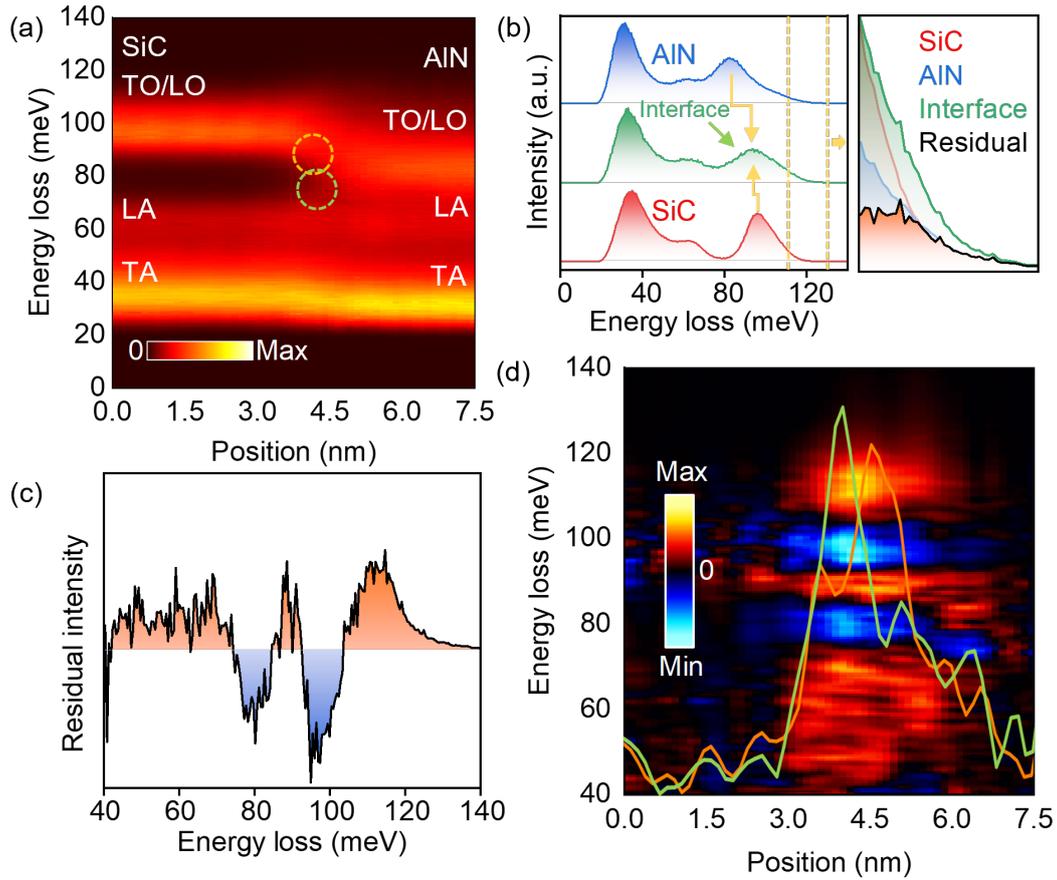

Figure 4. Nanometer-resolution phonon property measurements across the AlN-SiC interface. a) The phonon energy spectrum across the AlN-SiC interface, where the interfacial phonon modes are marked by colored circles. b) Phonon energy–intensity spectra from different sections across the interface, with the broadening and shifts of spectral peaks indicated. The right panel shows an enlarged view of the phonon signal between 110 and 130 meV, together with the corresponding residual spectrum. c) The interfacial residual spectrum, which cannot be expressed as a linear combination of the bulk phonon spectra; two important interfacial phonon modes are high-lighted. d) The linear-fitting residual spectrum mapping across the interface, along with the residual line profiles in two energy windows: 69-72 meV (light green) and 85-88 meV (orange).

The existence of such interfacial phonon modes contributes to energy exchange between branches (from SiC optical and acoustic modes to AlN optical modes) with distinct energies via inelastic scattering, thereby providing additional energy transfer pathways and resulting in a higher TBC.[29,43,44] The unique phonon modes in the interfacial region can also be expressed as the portion of the phonon spectra that cannot be represented as a linear combination of the bulk materials on the two sides.[45,46] The residual after subtracting the bulk contribution is shown in Figures 4b-c, the detailed fitting result is shown in Figure S2. This spectrum represents the phonon modes originating from the interface itself rather than from the bulk materials, further details can be found in the Methods section. The remaining residual is in good agreement with previous study.[47] At energies around 90 meV and 70 meV, clear positive residual peaks are observed, and they are marked by arrows in the corresponding colors. These residual modes are further classified as partially extended modes (PM),[42,48,49] as shown in Figure S3, where the AlN optical mode extends into the SiC side to promote additional energy exchange with the higher-energy SiC optical phonons or lower-energy SiC acoustic phonons.[47] Figure 4d shows the special distribution of the residual across the interface, indicating that most of the residual modes are localized near the interface and decay rapidly within ~2 nm, as shown in Figures S4 b-f. The integrated profiles of the two important energy windows mentioned above are also shown in Figure 4d. The residual spectra of the low-energy (69–72 meV) and high-energy (85–88 meV) modes are marked in their corresponding colors, and their FWHM values are approximately 2 nm. Although Figures 4c-d also reveal residual phonon modes in the energy ranges of

40–60 meV and 110–120 meV, these modes are not expected to contribute significantly to TBC. Phonons with energies below 60 meV can transmit directly across the interface through elastic scattering, limiting the contribution of additional transmission pathways. In contrast, the density of states (DOS) for phonons above 110 meV is negligible in both bulk materials, so no effective phonon transmission can occur in this high-energy range. These effective couplings and energy exchanges between bulk and interfacial phonons, together with the good matching of acoustic modes enabled by the high-quality interfacial structure, are the fundamental origin of the high TBC.[50]

To further understand the EELS results, MD simulations were performed. Figure 5a illustrates the atomic structure of the AlN-SiC interface used in MD simulations to investigate the thermal transport mechanism.[51] The corresponding vibrational DOS derived from MD is presented in Figure 5b (a more detailed, layer-resolved description is provided in Figure S8), juxtaposed with the EELS measurement results for comparison. The bulk spectra of SiC and AlN overlap significantly in the low-frequency acoustic region, supporting direct phonon transmission via elastic processes.[52] The vibrational DOS of AlN and SiC near the interface exhibits distinct features compared to their corresponding bulk DOS, which qualitatively aligns with the localized interfacial modes observed in the EELS measurements. It should be noted that the EELS intensity is not a direct measure of the DOS but is related to it through the electron-phonon interaction matrix elements.[53] Therefore, combined with the inherent limitations of the Tersoff interatomic potential, the MD simulation serves to provide a qualitative rationalization for the phonon features identified in the EELS data.

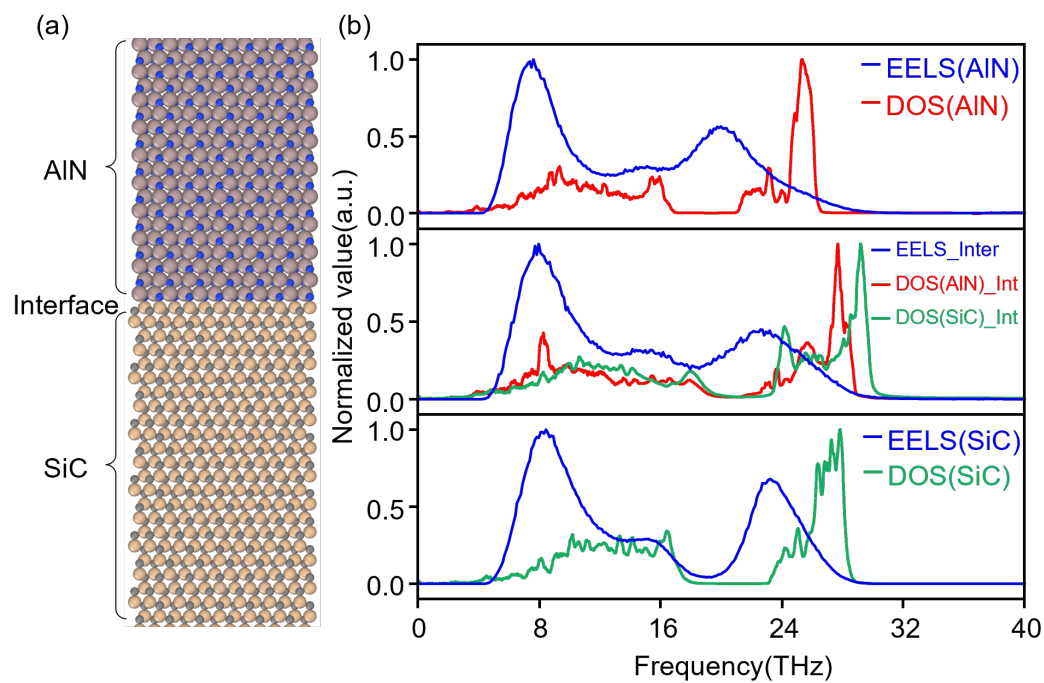

Figure 5. Calculation results of the MD simulation. (a) Atomic configuration of the AlN-SiC interface used in MD simulations. (b) Comparison of experimental EELS spectra against MD-calculated vibrational DOS for bulk AlN, the interface, and bulk SiC.

**Conclusions**

This work reported thermal transport across atomically sharp AlN-SiC interfaces, which achieves a record-high TBC of ~800 MW/m²-K—the highest value ever reported for this type of interfaces. This exceptionally high TBC arises from the high-quality AlN-SiC interface, fabricated via a novel ion implantation-induced nucleation epitaxial growth technique. TDTR measurements show that the thermal conductivity of the AlN films are close to the theoretical values. AGF calculations further reveal that the ultrahigh TBC is predominantly driven by elastic transmission channels, with nearly half of the acoustic modes exhibiting near-unity transmission. The integrated theoretical TBC (~735 MW/m²-K) accounts for over 90% of the experimental value, confirming the efficiency of these elastic pathways. Furthermore, utilizing high-energy-resolution EELS in STEM corroborated by MD simulations, unique interfacial phonon modes were identified to connect phonon spectra across the interface. These modes facilitate efficient inelastic phonon scattering and enhance heat transfer across the interfaces. This study underscores the vital role of interface structure in interfacial thermal transport and opens a route toward improved heat dissipation in high-power electronic devices.

**Methods**

**Materials growth**: An ion implantation induced nucleation (I³N) growth technique was employed to facilitate high-quality epitaxy. Specifically, the 4H-SiC substrates were subjected to Nitrogen (N) ion injection at room temperature, using an acceleration energy of 40 keV and a total dose of $1\times10^{12}$ cm$^{-2}$. Afterwards, a 430-nm-thick AlN film was synthesized via low-pressure metal-organic chemical vapor deposition (MOCVD). The growth conditions included a V/III ratio of 1500 and a high temperature of 1300°C. This method eliminated the conventional island-like morphology, resulting in an atomically abrupt interface. Following the deposition, selective etching was performed to achieve different thicknesses of 117 nm, 241 nm, and 337 nm on distinct segments, which is characterized by a profilometer (see Figure S5).

**Thermal property measurements**: Thermal characterization in this study is performed using TDTR, an ultrafast optical pump-probe technique for charactering the thermal properties of thin films and interfaces. Our experimental setup employs a mode-locked Ti: Sapphire laser source, generating 100-fs, 800 nm pulses at an 80 MHz repetition rate. This laser output is separated into two beams: the pump and the probe. The pump beam is first modulated at 5.02 MHz by an electro-optic modulator and then undergoes frequency doubling to 400 nm through a frequency doubling crystal (BiB$_3$O$_6$, BIBO) in a two-color configuration. Meanwhile, a motorized delay stage introduces a precisely controlled time shift to the probe beam, enabling picosecond-level temporal resolution. To guarantee complete spatial overlap, the pump beam is shaped to a slightly larger spot

(17.7 μm radius) than the probe (10 μm radius). The pump-induced temperature rise will induce a change in sample's reflectivity, which is detected by the probe beam. The signal is then demodulated with a lock-in amplifier to extract its in-phase (Vin) and out-of-phase (Vout) components. The thermal properties are ultimately determined by fitting the measured Vin/Vout ratio to a multi-layer heat transfer model. Besides, we use sensitivity coefficient $S_\alpha = \partial \ln R / \partial \ln \alpha$ to gauge the dependence of the measured ratio on specific parameter, which is presented in Figure S1 as an example. Picosecond acoustics technique is used to determine the film thickness, while its thermal conductivity was derived from four-point probe resistivity measurements using the Wiedemann-Franz law.

**TEM and EDS characterizations**: The sample used for STEM and EELS characterizations was prepared using a focused ion beam (FIB) system (Thermo Fisher Helios 5 UX) at Wintech-Nano (Suzhou) Co., Ltd. The atomic structure of the AlN layer and the AlN–SiC interfacial region was characterized using a spherical-aberration-corrected electron microscope (FEI Titan Cubed Themis G2 300). The EDS characterization of the sample was performed using an Arm200F.

**MD simulations and AGF calculations**: We investigate heat transport across the AlN-SiC interface shown in Figure 1a by combining MD and a mode-resolved AGF approach. Specifically, the interfacial vibrational DOS is extracted from MD, while phonon modal transmission is obtained using the mode-resolved AGF formalism.[35,51]

Interatomic interactions for Al–Al, Al–N, N–N, Si–Si, Si–C, and C–C are modeled with Tersoff potentials.[54,55] Owing to the minor lattice mismatch between AlN and SiC, identical unit-cell dimensions are adopted in the y and z directions (3.11 Å and 5.39 Å, respectively). Heat flow is evaluated along the $z$ direction, and periodic boundary conditions are imposed in the $x$ and $y$ directions (cross section).

Prior to AGF calculations, force constants are generated via lattice-dynamics calculations using the general utility lattice program (GULP).[56] These force constants are then used to assemble the harmonic matrix required by AGF. With translational invariance in the $x$ and $y$ directions, a Fourier transform is carried out on an 86 × 86 wave-vector mesh. The phonon spectral transmission $\Xi(\omega)$ is determined from the device retarded Green's function. To resolve transmissions by mode, the transmission matrix $t$ is constructed from Bloch matrices. The coupling between mode $i$ on one side and mode $j$ on the other side is expressed as $\Xi_{ij} = |t_{ij}|^2$. The modal transmission for mode $i$ is $\Xi_i(\omega) = \sum_j |t_{ij}|^2$, and the total spectral transmission follows $\Xi(\omega) = \sum_i \Xi_i(\omega)$. The thermal conductance per unit area $\sigma(T)$ of the interface can be calculated by the Landauer formalism: $\sigma(T) = \frac{1}{2\pi A} \int_0^{\omega_{max}} \hbar\omega \frac{\partial n}{\partial T}(\omega,T) \Xi(\omega) d\omega$, where A denotes the cross-sectional area, $n(\omega,T)$ is the Bose-Einstein distribution, and T is the temperature. According to this formula, larger modal transmission can lead to larger thermal conductance.

For vibrational DOS calculations, the cross section is chosen as 18.45 Å × 31.96 Å. MD

simulations are carried out using LAMMPS (large-scale atomic/molecular massively parallel simulator).[57] A time step of 0.5 fs is used at T = 300 K. The AlN and SiC segments are 79.72 Å and 80.56 Å long, respectively, so the total length of the system is 160.28 Å. The system is first equilibrated in the NPT ensemble for 0.5 ns, followed by 0.5 ns in NVE with a Langevin thermostat at 300 K, and then another 0.5 ns in NVE. finally, NVE ensemble is applied for another 0.5 ns. After equilibration, an additional 2.5 ns NVE run is performed while recording atomic velocities. The vibrational DOS is obtained by Fourier transforming the velocity autocorrelation function (VACF),[58] with $VACF(t) = \left\langle \sum_{i=1}^{N} v_i(t_0 + t)/N \right\rangle$, where $N$ is the number of atoms, $v_i(t_0)$ is the $i$th atom at time $t_0$, and $\langle \cdot \rangle$ represents averaging over a set of initial times $t_0$. Accordingly, $DOS(f) = \int_{-\infty}^{+\infty} VACF(t) e^{-2\pi i f t} dt$, where $f$ is the phonon frequency. Here, AlN_interface and SiC_interface layer refer to the first atomic layers adjacent to the interface, whereas bulk indicates regions sufficiently distant from the interface.

**High-energy-resolution EELS**: High-energy-resolution EELS spectra were acquired on a Nion U-HERMES200 microscope operated at 60 keV, with a convergence semi-angle of 35 mrad and a collection semi-angle of 25 mrad. The dataset was obtained over a mapping area of 60 × 11 pixels (8 × 1.5 nm), with an acquisition time of 1000 ms per pixel. The total acquisition time was approximately 10 minutes, with a sample drift of less than 1 nm, which was corrected during data processing. All EELS data were processed using a custom-written MATLAB code[46]. The position of the zero-loss peak (ZLP) at each point was first corrected by cross-correlation. A block-matching and 3D

(BM3D) method was then applied to each channel to remove Gaussian noise.[59] The ZLP was removed by fitting the acquired spectrum to a Pearson VII function in two energy windows, located before and after the phonon signal range. To reduce the broadening effect caused by finite energy resolution, Lucy–Richardson deconvolution with five iterations was applied, using the fitted ZLP as the point spread function. The interfacial residual mode $R(\omega)$ can be expressed as $R(\omega) = S_{int}(\omega) - a_1 S_{SiC}(\omega) - a_2 S_{AlN}(\omega)$. Where $S_{int}(\omega)$, $S_{SiC}(\omega)$, and $S_{AlN}(\omega)$ are the experimentally acquired spectra from the interfacial, SiC, and AlN regions, respectively. The coefficients $a_1$ and $a_2$ were obtained by minimizing $\| R(\omega) \|^2$ through linear least-squares fitting in MATLAB. The representative fitting result is shown in Figure S2. This method is commonly used to extract the pure interfacial vibrational modes.[45,46,60,61] The total squared norm of the residual and the corresponding fitting coefficients across the interface are shown in Figure S4b.

**Acknowledgements**

J.L., Z.H. and Z.C. thank Prof. Hui Li for the EDS measurements. Z.C. thanks Prof. Tianli Feng for helpful discussions. J.L., Z.H. and Z.C. acknowledge the funding support from the National Natural Science Foundation of China (Grants No. 62574007, T2550270), the National Key Research and Development Program of China (Grant No. 2024YFA1207900). The authors acknowledge Electron Microscopy Laboratory of Peking University, China for the use of Cs corrected Titan Cubed Themis G2 300 transmission electron microscopy and Cs corrected Nion U-HERMES200 scanning transmission electron microscopy.

## Competing interests

The authors declare no competing interest.

## Data availability

The data that support the findings of this study are available from the corresponding authors upon reasonable request.

# References

[1] Z. Cheng, J. Liang, K. Kawamura, H. Zhou, H. Asamura, H. Uratani, J. Tiwari, S. Graham, Y. Ohno, Y. Nagai, T. Feng, N. Shigekawa, D. G. Cahill, *Nat Commun* **2022**, *13*, 7201.

[2] T. L. Bougher, Yates ,Luke, Lo ,Chien-Fong, Johnson ,Wayne, Graham ,Samuel, B. A. and Cola, *Nanoscale and Microscale Thermophysical Engineering* **2016**, *20*, 22.

[3] Y. Zhang, F. Udrea, H. Wang, *Nat Electron* **2022**, *5*, 723.

[4] W.-Y. Woon, A. Kasperovich, J.-R. Wen, K. K. Hu, M. Malakoutian, J.-H. Jhang, S. Vaziri, I. Datye, C. C. Shih, J. F. Hsu, X. Y. Bao, Y. Wu, M. Nomura, S. Chowdhury, S. S. Liao, *Nat Rev Electr Eng* **2025**, *2*, 598.

[5] H. Zhou, C. Zhang, K. Zhang, Z. Huang, F. Liu, M. Zhou, H. Gong, S. Tang, W. Liu, B. Wang, Y. Dong, J. Liu, S. Zhou, Z. Xu, S. Wang, Z. Liu, S. Xu, C. Zhang, X. Wang, H. Wang, Y. Zhang, Z. Cheng, T. Chen, Y. Zhang, Y. Hao, J. Zhang, *Nat Commun* **2025**.

[6] Z. Cheng, Z. Huang, J. Sun, J. Wang, T. Feng, K. Ohnishi, J. Liang, H. Amano, R. Huang, *Applied Physics Reviews* **2024**, *11*.

[7] Z. Cheng, F. Mu, L. Yates, T. Suga, S. Graham, *ACS Appl. Mater. Interfaces* **2020**, *12*, 8376.

[8] S. Vaziri, C. Perez, I. M. Datye, H. Kwon, C. Hsu, M. E. Chen, M. Noshin, T. Lee, M. Asheghi, W. Woon, E. Pop, K. E. Goodson, S. S. Liao, X. Bao, *Adv Funct Materials* **2025**, *35*, 2402662.

[9] H. Walwil, Y. Song, D. C. Shoemaker, K. Kang, T. Mirabito, J. M. Redwing, S.

Choi, *Journal of Applied Physics* **2025**, *137*, 095105.

[10] R. Li, K. Hussain, M. E. Liao, K. Huynh, M. S. B. Hoque, S. Wyant, Y. R. Koh, Z. Xu, Y. Wang, D. P. Luccioni, Z. Cheng, J. Shi, E. Lee, S. Graham, A. Henry, P. E. Hopkins, M. S. Goorsky, M. A. Khan, T. Luo, *ACS Appl. Mater. Interfaces* **2024**, *16*, 8109.

[11] Z. Su, J. P. Freedman, J. H. Leach, E. A. Preble, R. F. Davis, J. A. Malen, *J. Appl. Phys.* **2013**, *113*, 213502.

[12] S. Tian, T. Wu, S. Hu, D. Ma, L. Zhang, *Applied Physics Letters* **2024**, *124*, 042202.

[13] T. Hwang, P.-C. Lee, A. C. Kummel, K. Cho, *ACS Appl. Mater. Interfaces* **2024**, *16*, 53098.

[14] M. Hu, P. Wang, D. Wang, Y. Wu, S. Mondal, D. Wang, E. Ahmadi, T. Ma, Z. Mi, *APL Materials* **2023**, *11*, 121111.

[15] X. F. Xinye Fan, Y. H. Yongqing Huang, X. R. Xiaomin Ren, X. D. Xiaofeng Duan, F. H. Fuquan Hu, Q. W. Qi Wang, *Chin. Opt. Lett.* **2012**, *10*, 110402.

[16] J. Lv, H. Huang, Y. Huang, X. Ren, A. Miao, Y. Li, H. Du, Q. Wang, S. Cai, *IEEE Transactions on Electron Devices* **2008**, *55*, 322.

[17] A. Iqbal, G. Walker, A. Iacopi, F. Mohd-Yasin, *Journal of Crystal Growth* **2016**, *440*, 76.

[18] W. Wei, Y. Peng, Y. Yang, K. Xiao, M. Maraj, J. Yang, Y. Wang, W. Sun, *Nanomaterials* **2022**, *12*, 3937.

[19] T. Metzger, R. Höpler, E. Born, O. Ambacher, M. Stutzmann, R. Stömmer, M. Schuster, H. Göbel, S. Christiansen, M. Albrecht, H. P. Strunk, *Philosophical Magazine*

*A* **1998**, *77*, 1013.

[20] Q. S. Paduano, A. J. Drehman, D. W. Weyburne, J. Kozlowski, J. Serafinczuk, J. Jasinski, Z. Liliental-Weber, *phys. stat. sol. (c)* **2003**, 2014.

[21] R. Xie, J. Tiwari, T. Feng, *J. Appl. Phys.* **2022**, *132*, 115108.

[22] Z. Cheng, Y. R. Koh, A. Mamun, J. Shi, T. Bai, K. Huynh, L. Yates, Z. Liu, R. Li, E. Lee, M. E. Liao, Y. Wang, H. M. Yu, M. Kushimoto, T. Luo, M. S. Goorsky, P. E. Hopkins, H. Amano, A. Khan, S. Graham, *Phys. Rev. Materials* **2020**, *4*, 044602.

[23] L. Lindsay, D. A. Broido, T. L. Reinecke, *Phys. Rev. B* **2013**, *87*, 165201.

[24] F. Mu, Z. Cheng, J. Shi, S. Shin, B. Xu, J. Shiomi, S. Graham, T. Suga, *ACS Appl. Mater. Interfaces* **2019**, *11*, 33428.

[25] J. Liang, H. Nagai, Z. Cheng, K. Kawamura, Y. Shimizu, Y. Ohno, Y. Sakaida, H. Uratani, H. Yoshida, Y. Nagai, N. Shigekawa, *Selective Direct Bonding of High Thermal Conductivity 3C-SiC Film to β-Ga2O3 for Top-Side Heat Extraction*, arXiv **2022**.

[26] J. S. Kang, M. Li, H. Wu, H. Nguyen, T. Aoki, Y. Hu, *Nat Electron* **2021**, *4*, 416.

[27] L. Yates, T. L. Bougher, T. Beechem, B. A. Cola, S. Graham, in *Volume 3: Advanced Fabrication and Manufacturing; Emerging Technology Frontiers; Energy, Health and Water- Applications of Nano-, Micro- and Mini-Scale Devices; MEMS and NEMS; Technology Update Talks; Thermal Management Using Micro Channels, Jets, Sprays*, American Society of Mechanical Engineers, San Francisco, California, USA **2015**, p. V003T04A009.

[28] M. Malakoutian, D. E. Field, N. J. Hines, S. Pasayat, S. Graham, M. Kuball, S.

Chowdhury, *ACS Appl. Mater. Interfaces* **2021**, *13*, 60553.

[29] J. T. Gaskins, G. Kotsonis, A. Giri, S. Ju, A. Rohskopf, Y. Wang, T. Bai, E. Sachet, C. T. Shelton, Z. Liu, Z. Cheng, B. M. Foley, S. Graham, T. Luo, A. Henry, M. S. Goorsky, J. Shiomi, J.-P. Maria, P. E. Hopkins, *Nano Lett.* **2018**, *18*, 7469.

[30] K. Woo, M. Malakoutian, Y. Jo, X. Zheng, T. Pfeifer, R. Mandia, T. Hwang, H. Aller, D. Field, A. Kasperovich, D. Saraswat, D. Smith, P. Hopkins, S. Graham, M. Kuball, K. Cho, S. Chowdhury, in *2023 International Electron Devices Meeting (IEDM)*, **2023**, pp. 1–4.

[31] M. Sakata, T. Oyake, J. Maire, M. Nomura, E. Higurashi, J. Shiomi, *Appl. Phys. Lett.* **2015**, *106*.

[32] Y. K. Koh, Y. Cao, D. G. Cahill, D. Jena, *Adv Funct Materials* **2009**, *19*, 610.

[33] C. Monachon, L. Weber, C. Dames, *Annual Review of Materials Research* **2016**, *46*, 433.

[34] Z.-Y. Ong, G. Zhang, *Phys. Rev. B* **2015**, *91*, 174302.

[35] L. Yang, B. Latour, A. J. Minnich, *Phys. Rev. B* **2018**, *97*, 205306.

[36] S. Tian, T. Wu, S. Hu, D. Ma, L. Zhang, *Applied Physics Letters* **2024**, *124*, 042202.

[37] S. Lyu, W. R. L. Lambrecht, *Phys. Rev. B* **2020**, *101*.

[38] C. Bungaro, K. Rapcewicz, J. Bernholc, *Phys. Rev. B* **2000**, *61*, 6720.

[39] Z. Cheng, Z. Huang, J. Sun, J. Wang, T. Feng, K. Ohnishi, J. Liang, H. Amano, R. Huang, *Applied Physics Reviews* **2024**, *11*, 041324.

[40] T. Feng, H. Zhou, Z. Cheng, L. S. Larkin, M. R. Neupane, *ACS Appl. Mater. Interfaces* **2023**, *15*, 29655.

[41] A. Giri, P. E. Hopkins, *Adv Funct Materials* **2020**, *30*, 1903857.

[42] Y.-H. Li, R.-S. Qi, R.-C. Shi, J.-N. Hu, Z.-T. Liu, Y.-W. Sun, M.-Q. Li, N. Li, C.-L. Song, L. Wang, Z.-B. Hao, Y. Luo, Q.-K. Xue, X.-C. Ma, P. Gao, *Proceedings of the National Academy of Sciences* **2022**, *119*, e2117027119.

[43] Q. Li, F. Liu, S. Hu, H. Song, S. Yang, H. Jiang, T. Wang, Y. K. Koh, C. Zhao, F. Kang, J. Wu, X. Gu, B. Sun, X. Wang, *Nat Commun* **2022**, *13*, 4901.

[44] P. E. Hopkins, J. C. Duda, P. M. Norris, *Journal of Heat Transfer* **2011**, *133*, 062401.

[45] Z. Cheng, R. Li, X. Yan, G. Jernigan, J. Shi, M. E. Liao, N. J. Hines, C. A. Gadre, J. C. Idrobo, E. Lee, K. D. Hobart, M. S. Goorsky, X. Pan, T. Luo, S. Graham, *Nat Commun* **2021**, *12*, 6901.

[46] R. Qi, R. Shi, Y. Li, Y. Sun, M. Wu, N. Li, J. Du, K. Liu, C. Chen, J. Chen, F. Wang, D. Yu, E.-G. Wang, P. Gao, *Nature* **2021**, *599*, 399.

[47] F. Liu, R. Mao, Z. Liu, J. Du, P. Gao, *Nature* **2025**, *642*, 941.

[48] M. Wu, R. Shi, R. Qi, Y. Li, T. Feng, B. Liu, J. Yan, X. Li, Z. Liu, T. Wang, T. Wei, Z. Liu, J. Du, J. Chen, P. Gao, *Chinese Phys. Lett.* **2023**, *40*, 036801.

[49] K. Gordiz, A. Henry, *Journal of Applied Physics* **2016**, *119*, 015101.

[50] S. Huang, F. Liu, R. Mao, Q. Guo, S. Li, F. Sun, Z. Wang, P. Gao, *ACS Appl. Mater. Interfaces* **2025**, acsami.5c00949.

[51] Z.-Y. Ong, G. Zhang, *Phys. Rev. B* **2015**, *91*, 174302.

[52] Z. Cheng, R. Li, X. Yan, G. Jernigan, J. Shi, M. E. Liao, N. J. Hines, C. A. Gadre, J. C. Idrobo, E. Lee, K. D. Hobart, M. S. Goorsky, X. Pan, T. Luo, S. Graham, *Nat*

*Commun* **2021**, *12*, 6901.

[53] R. Qi, N. Li, J. Du, R. Shi, Y. Huang, X. Yang, L. Liu, Z. Xu, Q. Dai, D. Yu, P. Gao, *Nat Commun* **2021**, *12*, 1179.

[54] M. Tungare, Y. Shi, N. Tripathi, P. Suvarna, F. (Shadi) Shahedipour-Sandvik, *physica status solidi (a)* **2011**, *208*, 1569.

[55] P. Erhart, K. Albe, *Phys. Rev. B* **2005**, *71*, 035211.

[56] J. D. Gale, A. L. Rohl, *Molecular Simulation* **2003**, *29*, 291.

[57] A. P. Thompson, H. M. Aktulga, R. Berger, D. S. Bolintineanu, W. M. Brown, P. S. Crozier, P. J. In 'T Veld, A. Kohlmeyer, S. G. Moore, T. D. Nguyen, R. Shan, M. J. Stevens, J. Tranchida, C. Trott, S. J. Plimpton, *Computer Physics Communications* **2022**, *271*, 108171.

[58] G. S. Grest, S. R. Nagel, A. Rahman, T. A. Witten, *The Journal of Chemical Physics* **1981**, *74*, 3532.

[59] K. Dabov, A. Foi, V. Katkovnik, K. Egiazarian, *IEEE Trans. on Image Process.* **2007**, *16*, 2080.

[60] S. Huang, F. Liu, J. Liu, X. Gao, Z. Wang, P. Gao, *Adv Materials Inter* **2025**, *12*, 2400816.

[61] S. Huang, F. Liu, J. Liu, R. Mao, J. Zhang, Z. Liu, F. Sun, Z. Wang, P. Gao, *ACS Appl. Mater. Interfaces* **2025**, *17*, 46409.